\documentclass[10pt,twocolumn]{article}
\usepackage[top=1.9cm, bottom=1.9cm, left=1.8cm, right=1.8cm]{geometry}
\usepackage{times}  %
\usepackage[sort&compress,numbers]{natbib}  %
\usepackage[hyphens]{url}
\usepackage{graphicx}  %
\usepackage[keeplastbox]{flushend}
\usepackage[hyphens]{url}  %
\usepackage{graphicx} %
\usepackage{tikzsymbols}

\newcommand{\descr}[1]{\smallskip\noindent\textbf{#1}}

\usepackage{sectsty}
\sectionfont{\bfseries\Large\raggedright}

\usepackage{url}                                %
\usepackage{array,multirow,graphicx,adjustbox}  %
\usepackage{booktabs}                           %
\usepackage[utf8]{inputenc}
\usepackage[ruled,algosection,noend,linesnumbered]{algorithm2e}
\usepackage{float}
\usepackage{paralist}
\usepackage{amsmath}
\usepackage{amsfonts}
\usepackage{tcolorbox}
\usepackage{xcolor}
\usepackage{csquotes} 
\usepackage{tabularx}
\usepackage{makecell}
\usepackage{pbox}
\usepackage[hang,flushmargin]{footmisc}
\usepackage{flushend}
\usepackage{xspace}
\usepackage{multicol, lipsum}
\usepackage[T1]{fontenc}

\usepackage{xcolor}
\usepackage{booktabs}
\usepackage{graphicx}
\usepackage{paralist}
\usepackage[small,bf]{caption}
\usepackage{subcaption}

\let\oldbibliography\thebibliography
\renewcommand{\thebibliography}[1]{%
  \oldbibliography{#1}%
  \setlength{\itemsep}{2pt}%
}

\usepackage[hang,flushmargin]{footmisc}

\usepackage[compact]{titlesec}
\titlespacing*{\section}{0pt}{*3}{3pt}
\titlespacing*{\subsection}{0pt}{*2}{2pt}
\usepackage{xspace}

\makeatletter
\def\url@leostyle{%
  \@ifundefined{selectfont}{\def\UrlFont{}}%
  {\def\UrlFont{}}%
}
\makeatother
\usepackage[hyphenbreaks]{breakurl}

\usepackage[bookmarks=true, bookmarksnumbered=true, colorlinks=true, linkcolor=linkcol, citecolor=citecol, urlcolor=urlcol, hypertexnames=true]{hyperref}

\definecolor{darkgreen}{RGB}{0, 100, 0}
\definecolor{linkcol}{rgb}{0.3,0,0}
\definecolor{citecol}{rgb}{0.3,0,0}
\definecolor{urlcol}{rgb}{0.3,0,0}

\makeatletter
\def\url@leostyle{%
  \@ifundefined{selectfont}{\def\UrlFont{\small}}%
  {\def\UrlFont{}}%
}
\makeatother
\newif\ifcomment
\commenttrue
\ifcomment
\newcommand{\sz}[1]{{\bf \textcolor{brown}{SZ: #1}}}

\newcommand{\cc}[1]{{\bf\textcolor{orange}{CC: #1}}}
\else
\newcommand{\sz}[1]{}
\newcommand{\cc}[1]{}
\fi

\title{\bf Slapping Cats, Bopping Heads, and Oreo Shakes: Understanding Indicators of Virality in TikTok Short Videos}

\begin{document}
\sloppy

\author{Chen Ling$^1$, Jeremy Blackburn$^2$, Emiliano De Cristofaro$^3$, and Gianluca Stringhini$^1$\\[0.5ex]
\normalsize{$^1$Boston University, $^2$Binghamton University, $^3$University College London}\\
}
\date{}

\maketitle

\begin{abstract}
Short videos have become one of the leading media used by younger generations to express themselves online and thus a driving force in shaping online culture.
In this context, TikTok has emerged as a platform where  viral videos are often posted first.
In this paper, we study what elements of short videos posted on TikTok contribute to their virality. 
We apply a mixed-method approach to develop a codebook and identify important virality features.
We do so vis-\`a-vis three research hypotheses; namely, that: 1) the video content, 2) TikTok's recommendation algorithm, and 3) the popularity of the video creator contribute to virality.

We collect and label a dataset of 400 TikTok videos and train classifiers to help us identify the features that influence virality the most.
While the number of followers is the most powerful predictor, close-up and medium-shot scales also play an essential role.
So does the lifespan of the video, the presence of text, and the point of view.
Our research highlights the characteristics that distinguish viral from non-viral TikTok videos, laying the groundwork for developing additional approaches to create more engaging online content and proactively identify possibly risky content that is likely to reach a large audience.
\end{abstract}

\section{Introduction}
\label{sec:intro}

The ubiquity of mobile devices and their increased bandwidth allow larger and larger portions of the population to upload and stream videos anywhere, any time.
This also enables short video platforms to become popular, especially among young adults.
One such platform is TikTok, a social media app developed in China that allows users to upload short (up to one minute) videos. 
By early 2021, TikTok attracted 689 million users, with a global penetration estimated at 18\% of global Internet users aged 16--64~\cite{sun2020content}.

Short video sharing platforms provide a completely different user experience than other kinds of social apps, as they no longer require prolonged attention from the viewer.
They also allow content creators to express themselves more directly, leaving a stronger impression on the viewer.
Consequently, social video-sharing apps have formed a novel ecosystem and have started shaping popular culture.
According to TikTok~\cite{tiktok2020viralvideo}, the top 10 most popular videos of 2020 attracted over 120 million {\em likes.}
These videos feature the creator acting in front of the camera and/or sharing random moments from their life.
 For example,
a video showing a man skateboarding home in the sunset attracted 12.7M {\em likes},
while a video making impressions of ``that coworker you love to hate'' during Zoom meetings, received 1M {\em likes} and amassed 2M followers to the account on TikTok.
Despite the importance that viral short videos have in shaping popular culture, the research community does not have a good understanding of what elements in a short video may contribute to it going viral.

In this paper, we present the first attempt to fill this gap.
We do so vis-\`a-vis three research hypotheses:

\begin{enumerate}
\item[(RH1)] {\em Content Elements.} Video content has an impact on the virality of short videos.
For example, TikTok's famous dancing challenge attracts millions of users participating in this trend by recording a clip of their movement with music. 
Users are also attracted by videos featuring kittens and puppies~\cite{o2014cats}.
\item[(RH2)] {\em Recommendation System.} Adding a trending hashtag in the video description helps short videos go viral.
  In other words, ``exploiting'' TikTok's recommendation algorithm might help make videos go viral.
  (Generic hashtags, including \#fyp and \#foryou, are added to video descriptions with the goal of featuring in the recommendation video stream.)
\item[(RH3)] {\em Creator's Profile.} The status of the creator affects the chances that a short video will go viral. 
TikTok provides verified badges for users as a blue checkmark on their profile page~\cite{tiktok2019verified}.
Popular creators on TikTok are also gaining extra visibility as they attract millions of followers.
\end{enumerate}

\descr{Methods.}
We use a mixed-methods strategy. 
First, we gather videos (along with metadata) featuring popular hashtags on TikTok. %
These videos serve as a starting point to identify indicators that may impact a video's virality. 
More specifically, we consider the number of {\em likes} on a video as a measure of its virality.
Then, we develop a codebook to characterize the indicators of virality along with our three research hypotheses and label 400 videos according to them.
Finally, we train a classifier on this set of 400 videos to distinguish between viral and non-viral videos. 

\descr{Results.}
Our analysis yields several interesting results:
\begin{enumerate}
  \item We build machine learning models based on the features from our codebook and show that they can effectively distinguish between a short video that will go viral and one that will not. The best classifier, Logistic Regression, yields a 0.93 Area Under the ROC Curve (AUC).
  \item  Our classifiers allow us to identify the most important features that discriminate between viral and non-viral videos. 
  The creator's popularity has the most substantial influence on the video's virality (RH3), while the scale and point of view of a video also play an essential role for virality (RH1); in particular, a close-up or medium-scale in a second-person view video helps in making videos go viral. 
  In contrast to previous studies on image memes, we find that including text captions in videos helps with their virality (RH1).
  In addition, a video posted including a hashtag while that hashtag is trending has a higher chance to go viral (RH2). %

  \item Our model generalizes beyond the training dataset; our classifiers help characterize short TikTok videos that are posted on Twitter, as well as those published under the popular hashtag \#fyp on TikTok. 
  Classifiers trained on the TikTok platform's trending short videos data correctly predict the virality of random TikTok short videos shared on Twitter and the most viral videos on TikTok.
  \item We provide examples to show that our features can help explain why specific short videos become popular while others experience a minimal distribution.
\end{enumerate}

\section{Background \& Related Work}
\label{sec:relatedwork}

In this section, we introduce TikTok's affordances and how they influence how people use the platforms.
We then review past research on the virality of online content.
Finally, we discuss related work on measuring and characterizing videos posted online.

\subsection{TikTok}

\begin{figure*}[t]
\centering
\begin{subfigure}[b]{.325\textwidth}
\centering
\includegraphics[width=0.65\linewidth]{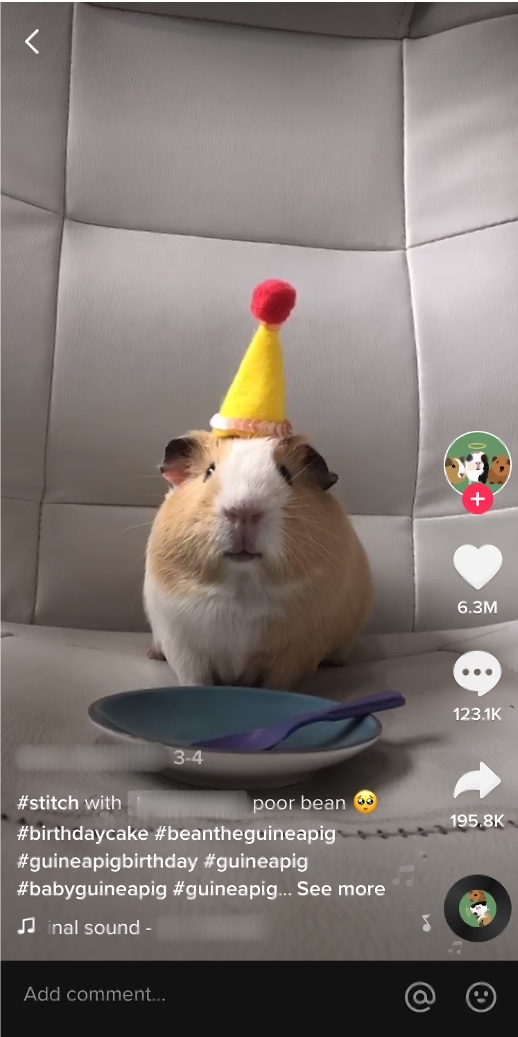}
\caption{{\em Video} page.}
\label{fig:videopage}
\end{subfigure}
\begin{subfigure}[b]{.325\textwidth}
\centering
\includegraphics[width=0.65\linewidth]{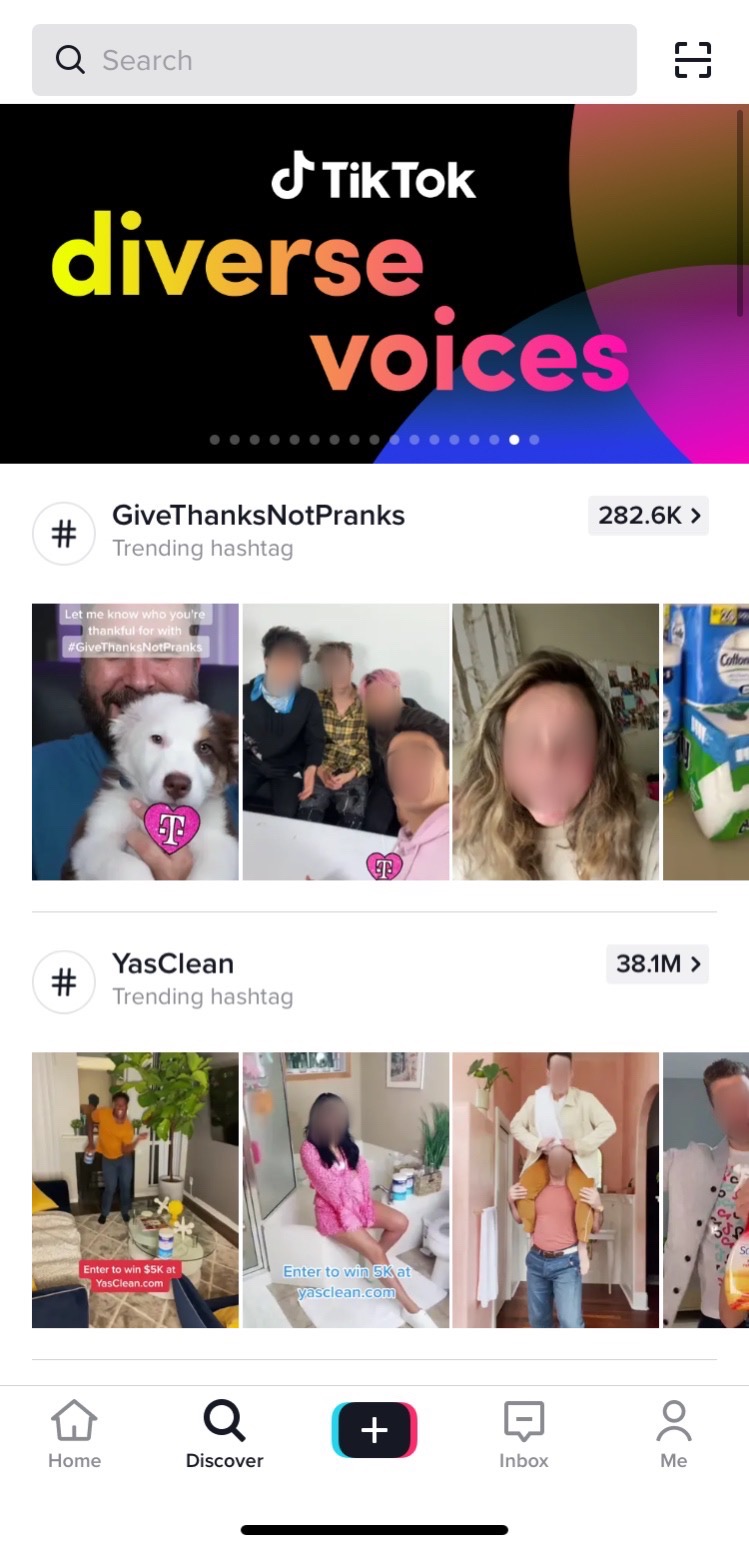}
\caption{{\em Discover} page.}\label{fig:discoverpage}
\end{subfigure}
\begin{subfigure}[b]{.325\textwidth}
\centering
\includegraphics[width=0.65\linewidth]{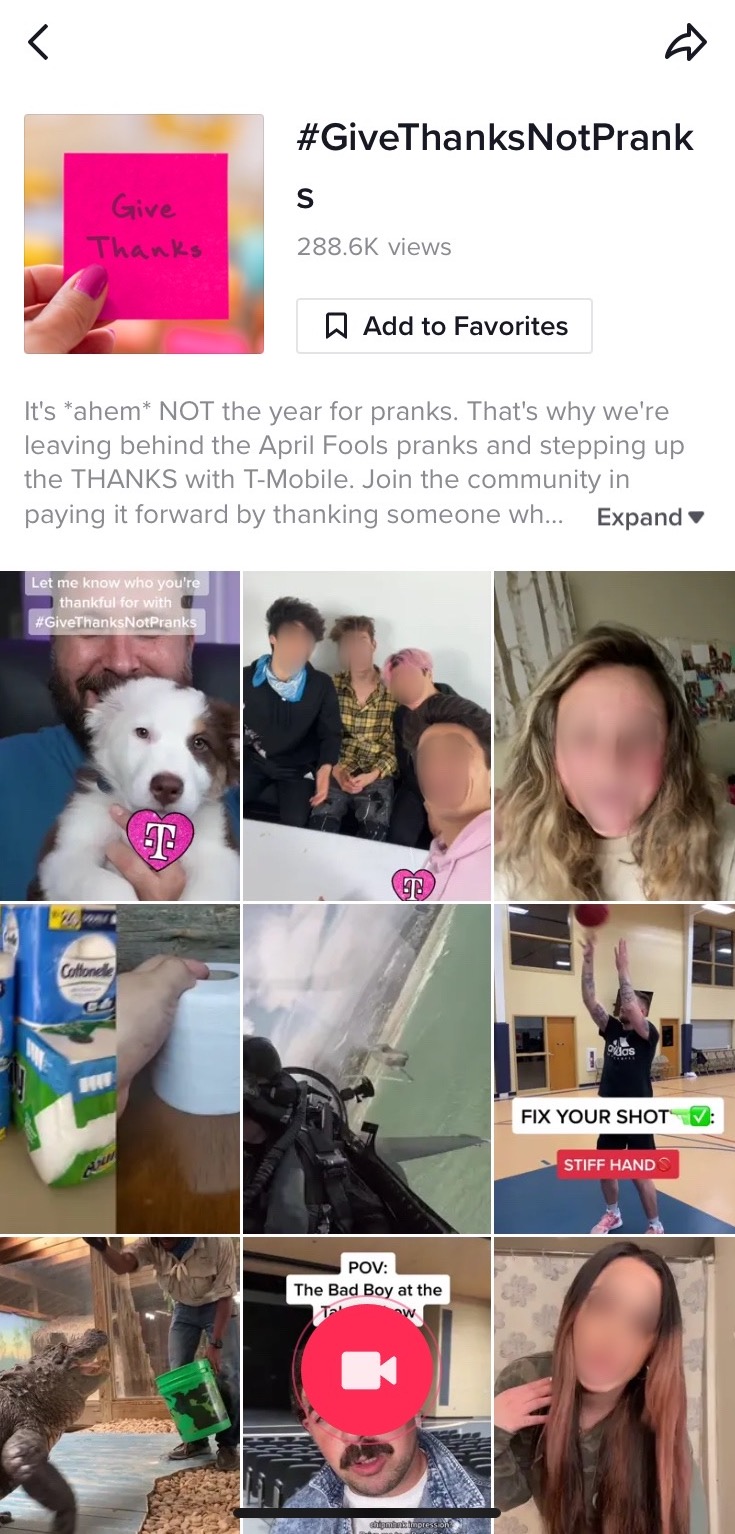}
\caption{A hashtag search result page.}
\label{fig:eventpage}
\end{subfigure}
\caption{Interface of TikTok app. Faces are blurred for privacy reasons.}
\end{figure*}

Unlike first-generation social media, TikTok is designed to be mobile-centered. 
Although users can view videos using a browser, functions are limited, as comments are not accessible, and users cannot create a video through a non-mobile device.

TikTok's user interface design is possibly the main driver in increasing user engagement~\cite{zulli2020extending}.
When opening the app, users are taken to the {\em Home} page, where algorithmically selected -- {\em ``For You''} -- videos play automatically; see Figure~\ref{fig:videopage}.
These are personalized feeds, and past viewing behavior influences the content displayed on the {\em Home} page.
The user can then switch to the {\em Following} Tab to watch feeds from users that they actively follow.

Like other online social services, TikTok encourages creators to use hashtags to describe, categorize, and make their videos easy to find.
The most popular hashtags over time represent the platform's most popular topics, which are displayed on the {\em Discover} page; see Figure~\ref{fig:discoverpage}.
On the {\em Discover} page, users can also search for hashtags and sounds. 
This page is the same for all Tiktok users and is unaffected by their activity on the platform.

Memetic remixes are also encouraged on TikTok, which often become viral.
We use a consistent meme definition in images and text as a piece of culture, typically with sarcastic or amusing undertones, which gains influence by being repeated, modified, and shared online~\cite{davison2012language, ling2021dissecting}.
In less than a year, the most popular videos on TikTok have received over 50 million {\em likes.}
For instance, a 15-second video where a girl just bops her head to Millie B's ``M to the B'' song, turning it into a TikTok sensation.
This song has since been used in more than 5 million videos in a memetic manner.

With its rapid global expansion, TikTok elicits more than just entertainment.
On the one hand, TikTok is dedicated to promoting positive outcomes.
For instance,  during the COVID-19 pandemic, they encouraged the use of masks and provided users with authoritative information and guidelines~\cite{eghtesadi2020facebook}; they also teamed up with scientists to create the hashtags \#scienceathome and \#learnonTikTok and promote the platform's educational benefits~\cite{zeng2020reposting}.
On the other hand, TikTok has often been criticized over privacy~\cite{neyaz2020security}, hate speech~\cite{weimann2020research}, censorship~\cite{melin2021china}, and cyberbullying~\cite{anderson2020getting} issues.
Inappropriate marketing strategies are also allegedly used in promoting e-cigarettes~\cite{delnevo2020rapid}, e.g., by portraying underage youth using them~\cite{tan2020puffbar}.

\subsection{Virality}

Marketing and psychology studies find that entertainment, socializing, convenience, status, information seeking, and structure are why people like online content~\cite{ahlse2020s}.
Affiliation, which refers to the human need to belong and form relationships, may promote the sharing of online content~\cite{baumeister1995need}.
People share different forms of content on social media to encourage others to connect with them and interact with others~\cite{ellison2007benefits}.
Research in psychology finds that a high physiological arousal status, i.e., excitement, panic, etc., makes people more likely to share online information~\cite{berger2011arousal}.
Marketing studies make a similar finding that virality is partially driven by physiological arousal~\cite{berger2012makes}, and that emotions generated by the content of advertisements influence its virality~\cite{nelson2013emotions}.

Researchers find that the characteristics of online virality depend on the type of online content.
For textual content, previous studies show that the success of online textual content depends on timing, the social network structure of the account posting it, randomness, and many other factors~\cite{centola2010spread,weng2012competition}.
Other research finds that textual content might become viral simply because it evokes a higher physiological arousal~\cite{berger2012makes}.

For image content, visual elements, including their aesthetic composition and catchy subjects, impact their virality.
A close-up of a character's facial expression attracts more attention than those larger-scale images with other subjects~\cite{ling2021dissecting}.

For video content, previous research finds that the popularity of Twitch live streamers improves the prediction of relative growth in viewers~\cite{netzorg2018popfactor}.
Studies on YouTube find that the recommendation algorithm~\cite{wu2019estimating} as well as in-links and fan base~\cite{khan2014virality} contribute to a video's popularity. 
Broxton et al.~\cite{broxton2013catching} show that a higher number of shares does not imply a higher number of views. %

Rizoiu et al.~\cite{rizoiu2017online} use Hawkes Intensity Process (HIP) models to assess videos under promotion in terms of their popularity growth per unit of promotion and the time it takes to launch such effects. 
Another study~\cite{vallet2015characterizing} finds a link between a high number of views on YouTube and a rapid spread of the video link on Twitter, demonstrating that user engagement in third-party communities has excellent cross-platform prediction capabilities. 
Researchers also investigate YouTube video view count as an indicator of virality from an epidemiological standpoint, demonstrating that attention dynamics follow similar patterns as infection rates~\cite{bauckhage2015viral}.
Finally, Wu et al.~\cite{wu2018beyond} find that the average fraction of a Youtube video watched by users is more stable over time than popularity, which varies over time and is influenced by external promotions.

In this paper, we use the number of likes that a video receives compared to all other videos containing the same hashtag as a proxy for its virality.
This is mainly because other potential numbers, including the number of shares and views, are not available for every video.
In addition, TikTok's swiping design for switching between videos may cause a bias in the number of views, as videos start playing automatically, but users might not watch the video completely.

\subsection{Understanding Video Content}

Previous work has studied video content along multiple axes.
Research on YouTube has studied its effectiveness as an educational resource for teachers and students~\cite{sherer2011using}.
Online video education achieves positive outcomes not only limited to traditional courses, e.g., STEM teaching~\cite{otchie2020can}, music~\cite{rudolph2009youtube}, physical education~\cite{quennerstedt2013pe}, but also for teenager-specific educations, like sex~\cite{johnston2017subscribing} and drug education~\cite{manning2013youtube}, as well as professional training, including anatomy~\cite{jaffar2012youtube}, nursing training~\cite{logan2012using}, etc.

Another line of work has focused on the advertisements implemented in online videos.
Previous studies on marketing confirm the prevalence and effectiveness of promotion for products like beauty, fashion~\cite{schwemmer2018social}, tourism~\cite{reino2011use}, and higher education~\cite{pham2017using}.
Research on YouTube Kids shows that advertising is disguised as other content to reach that target audience as one of the marketing strategies~\cite{araujo2017characterizing}.

In addition to exploring how online videos are used to pursue  educational progress and commercial interests, researchers have examined the social issues and political impact of online videos.
Political campaigns migrate from traditional television spots to online video platforms.
Prior work suggests that, although the same audience watches online and offline political advertisements,  online video advertisements are more likely to be shared~\cite{borah2018television}.

Researchers have studied the use of online videos to promote public health.
For example, videos on YouTube represent a vast source to educate the viewer to eat healthier with different emphasize~\cite{zhang2017persuading}.
Studies on vaccination decision-making show that most videos hold a positive stance, but anti-vaccine videos are the most liked and shared~\cite{covolo2017arguments}.

Prior work has also focused on malicious activities by analyzing user and video features to detect content that promotes hate, extremism~\cite{agarwal2014focused}, violence~\cite{giannakopoulos2010multimodal}, etc., or violating privacy~\cite{aggarwal2014mining}.

\section{Dataset}
\label{sec:dataset}

As mentioned, users find interesting videos on the {\em Discover} page; this allows them to search and explore the vast amount of content available on TikTok, and more specifically, popular videos, hashtags, creators, and sponsored content~\cite{tiktok2021discover}.

We visit the {\em Discover} page daily, between March 11 and April 11, 2021, and collect the 63 hashtags that appear there.
As TikTok allows searching for videos with a specific hashtag and viewing the most popular results according to its algorithm, we then collect a list of videos that appear in searches for these 63 hashtags.
The hashtag search returns a limited number of videos.
This seems to depend on the actual query, but most queries generate around 1,800 videos (it is not clear how this limit is defined~\cite{medina2020dancing}).
The popularity of the videos may play a role in the stream of videos resulting from the hashtag search; see Figure~\ref{fig:eventpage}.
The popularity of the hashtags may also impact the number of videos that we are able to retrieve, as TikTok only returns a fixed number of videos per query. %
The collection process produces 103,587 videos.

As we are interested in finding hashtags that contain both viral and non-viral videos, we remove those that only contain popular videos with 100 likes or more. %
This yields a dataset of 28,342 videos from 20 hashtags.

Following the same approach as previous work on characterizing virality~\cite{ling2021dissecting}, we label the 10 videos that attract the most {\em likes} from a hashtag as {\em viral}, and the 10 videos that receive the least {\em likes} from a hashtag as {\em non-viral.}
In total, we consider 400 videos for further analysis.
These videos serve as a basis to study our research hypotheses.

\descr{Ethical and Privacy Considerations.} 
Our study only uses publicly available information and entails no interaction with human subjects.
Therefore, this work is not considered human subjects research by the IRB at our institution.
Nonetheless, there are important ethical considerations to be made when analyzing social media data.
For instance, as human faces appear in our dataset, there might be risks to TikTok users, e.g., attracting unwanted attention. 
To mitigate this, we blur faces in all video examples. %
We also apply standard ethical guidelines, including only reporting data in the aggregate and not attempting to deanonymize online users~\cite{kenneally2012menlo}.

\section{Codebook}
\label{sec:characterizing}

In this section, we develop a codebook to guide the annotation process for our datasets with the goal of understanding indicators to make a short video go viral.
We follow a similar methodology as done in recent research on image meme virality~\cite{ling2021dissecting}.
More precisely, we followed three steps:
1) Three researchers independently screened our dataset and produce initial codes using thematic coding~\cite{braun2006using}.
2) We then discussed these initial codes using a subset of the dataset and agreed on the codebook by comparing agreement, see~ Tabel\ref{tbl:agreement}.
3) The first author annotated the rest of our dataset.

Inspired by previous research and by the affordances of the TikTok platform (see Section~\ref{sec:relatedwork}), we propose that viral videos should follow similar aesthetic properties as viral image memes~\cite{ling2021dissecting}.
At the same time, we consider the effect of the recommendation system and of the popularity of the creator on the virality of a video.
We identify a number of elements ({\em Features}) that are potentially characteristic of short video's virality and that can help us answer our three research hypotheses:

\descr{{\em RH1 Content Elements:} }
Our first hypothesis is that the elements of the video content have an effect on the likelihood of its virality.
Research in human vision suggests that the viewer's attention tends to be attracted by faces~\cite{buswell1935people}.
In particular, by faces of characters on a close-up scale with less text~\cite{ling2021dissecting}.
Emotions are also found to increase users' engagement in watching a video~\cite{lewinski2014predicting}.
In addition, short videos in first-person perspective and various styles (e.g., focusing on characters in medium or small shots, conveying positive or negative emotions, etc.) may receive more likes.

\descr{{\em RH2 Recommendation System:} }
We argue that users frequently try to promote their videos by utilizing TikTok's recommendation system. 
To this end, creators publish their videos including the hashtags that are trending on the {\em Discover} Page. %
We also find that creators sometimes also add additional, {\em unrelated} hashtags to promote their videos.

\descr{{\em RH3 Creator Profile:}}
Finally, the creator's profile may influence their videos' virality.
TikTok provides a verified badge for users~\cite{tiktok2019verified}, and popular creators have obviously more followers than average users.
We hypothesize that the creator's relationship with the platform impacts their video's virality.

\subsection{RH 1: Content Elements}
We posit that, to some extent, the content of a short video determines its virality~\cite{dupuis2019spread,crovitz2020analyzing,ling2021dissecting}.
Viewers may be more engaged when special themes are featured, i.e., disabled people thriving to shine, gorgeous pups roaming around the park, or babies gazing at their parents. 
Additionally, when making a short video, the way the content is presented, as well as the point of view, are all critical considerations.

\descr{F.1 Type of subject.} 
The types of subjects that can be depicted in a video include scenes, people, animals, or objects.
A new movie genre, \emph{desktop film}, which uses screen recording to treat the computer screen as both a camera lens and a canvas, is the other popular format in short videos~\cite{wiki2010computerfilm}, too.
TikTok creators record their screen to share live gaming, news podcasts, or their digital life.
Looking at the 400 videos in our dataset, we identify four types of subjects appearing in them: object, character, scenes, and screen recording. 
More precisely, we characterize subjects as follows:

\begin{itemize}
  \item ``Object'' refers to a material thing that can be seen and touched, like a table, a bottle, a building, or even a celestial body.
  \item ``Character'' refers to people, animals, or anthropomorphized objects, such as cartoon characters. 
  \item We categorize a subject as ``scene'' when the situation or activity depicted in a video is its main focus, instead of it being on the single characters or objects depicted in it.
  \item ``Screen recording'' includes screen captures of video games, news, TV shows, etc~\cite{opfer2011development}. 
\end{itemize}

Previous research showed that including special characters in online content elicits empathy and engagement in the viewers~\cite{alloway2014facebook,santoso2019social, o2014cats}.
As such, we identify the following features representing the characters in a video:

\begin{itemize}
\item Young child and infants, who have not matured sufficiently to speak as fluently as creators.
\item Person with a disability, i.e., people who have a physical disability, autism spectrum disorder, intellectual disability, mental health conditions, etc. 
\item Pet, i.e., puppies, kittens, other animals appear as a character in the video.
\end{itemize}

\descr{F.2 Scale.}  
Research in human vision showed that the viewer's gaze is biased towards the center of a scene (i.e., center bias)~\cite{buswell1935people,parkhurst2003scene,tatler2007central}.
We hypothesize that a video that is a close up of a subject will facilitate viewer focus on the salient part of the video and therefore catch their attention, while a large scale scene in which it is hard to identify the part to focus on might fail in attracting the viewer's attention.
To investigate how these aspects might affect the virality of a video, we consider its scale, which takes into account how the main subject is put in relation to the layout of the remaining elements of the video.
Most short videos are consistent in scale.
Based on the definitions of shots used in film studies~\cite{arijon1991grammar,tsingalis2012svm}, we define three scales for videos: close up, medium shot, and long shot.

\descr{F.3 Point of View.}
A first-person perspective provides a more immersive experience while also increasing social presence and entertainment value~\cite{maredia2018can}. 
First-person view videos on a short video platform, on the other hand, may have shaking cameras, uninteresting, or out-of-focus subjects.
By facing directly towards the audience, a second-person view is a more stable and convenient way to create high-quality content as well as promote the personal brand. 
We consider the following points of view in our codebook:

\begin{itemize}
\item First-person view:  a representation of what the creator sees~\cite{taylor2002video}.
For one of the popular subjects on TikTok: pets, we regard most of the videos shot by its owner and label them as the first-person view, unless an explicit clue of other viewpoints, for example, facing the camera with its owner, or being recorded by CCTV.

\item Second-person view: the character is talking to the viewer in front of the camera~\cite{ng2020you2me}.
TikTok videos are mostly created by a second-person view, with the creator talking or dancing in front of the audience back on the camera.

\item Third-person view: The story about ``them.''
The viewer has a limited point of view presents the action from the perspective of an ideal observer~\cite{taylor2002video}.
\end{itemize}

\descr{F.4 Text.} 
TikTok allows creators to add text on video for guiding viewers.
Some creators also add text to attract more followers.
We posit that the text on the video helps better illustrate its idea.

\descr{F.5 Emotion Experience.} Humorous and persuasive messages can significantly increase the sense of presence and subsequently facilitate message recall~\cite{skalski2009effects,wang2020humor}. 
Past research showed that the effectiveness of amusing video advertisements relates to people's happiness~\cite{lewinski2014predicting}.
We propose that videos elicit a stronger emotional response, increasing user engagement and causing them to {\em likes} the video.

\descr{F.6 Style.} 
TikTok is known for videos that include lip-syncing, dance routines, comedy skits, life-sharing shows, and make-up tutorials~\cite{hayes2020making}.
There are different styles
for users to create a video with a certain hashtag.
style types are listed but not limited as follows:
\begin{itemize}
\item Duets: duets enable users to remix other people's videos. 
Someone may, for example, upload a video of them swinging their arms around. 
Other TikTok users can then take that video and add themselves executing a similar activity.

\item Cringe: cringe refers to someone is performing awkwardly or embarrassingly despite striving to perform seriously. The creator takes the most embarrassing videos and makes reaction videos out of them.

\item Challenges: challenges refers to the creators who create films aiming to perform the same thing. 
For example, dancing challenges.
\end{itemize}

The same style creates memetic effects by linking videos together and allowing them to cross over with one another~\cite{tiktok2019guide}.
We believe that the creator's usage of a different style from the others may harm the virality of the video.

\subsection{RH 2: Recommendation System} 

We suggest that video creators leverage the recommendation system to make their content go viral.
Creators choose the appropriate time to publish their videos.
More specifically, creators use hashtags that are currently trending or have proven popular in the category~\cite{knowledge2019TikTok}.
We go over each feature regarding this RH in detail.

\begin{figure}[t]
\centering
\includegraphics[width=0.7\linewidth]{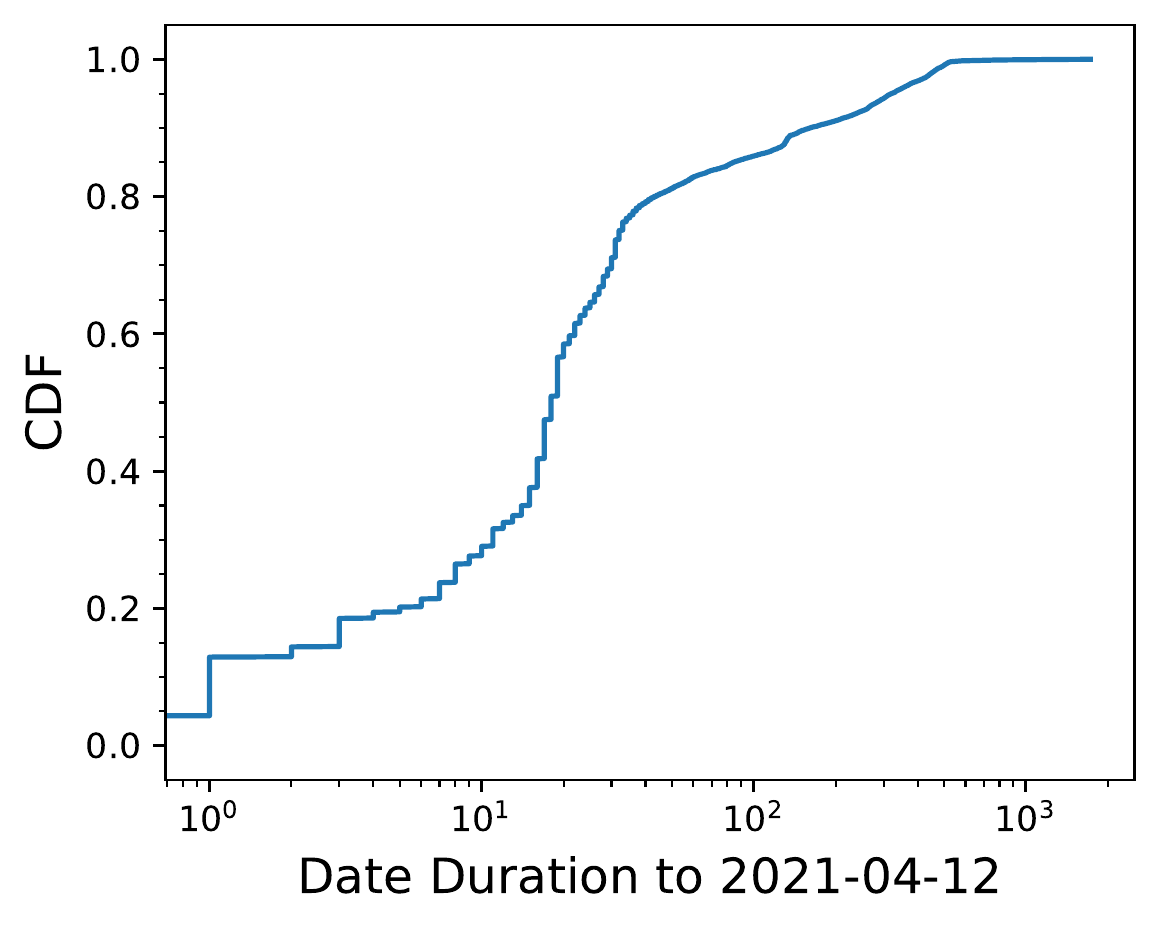}
\caption{Distribution of lifespan of the videos.}
\label{cdf:days}
\end{figure}

\descr{F.7 Lifespan.}
Lifespan refers to the time that has passed since a video was posted on TikTok.
Studies in viral information propagation indicate that online events that go viral are likely to do so shortly after they are posted~\cite{freeman2014viral}.

We suggest the videos created with the majority of other videos originated from the same hashtag search can take advantage of promotion of the hashtag on the {\em Discover} page.
From the distribution of lifespan on TikTok, see Figure~\ref{cdf:days}, we find 85\% of videos in our dataset are created within less than 100 days.
We propose videos posted from January 1 until April 11, 2021, are at the prime of their lifespan and can take advantage of video promotion on TikTok.

\descr{F.8 Hashtag.}
On TikTok, hashtags are crucial to users to locate videos and join conversations they are
interested in.
From the analysis of hashtag frequency (see Table~\ref{tbl:termfrequency}), we find there are two types of hashtags that are not directly related to videos: 1) generic hashtags that are believed to influence the recommendation algorithm, and 2) unrelated hashtags that are added to exploit trending topics and make the video go viral.

The first type are the most popular generic hashtags on TikTok, including ``foryou,'' ``fyp,'' ``xyzbca,'' etc.~\cite{mau2021popularhashtags, tiktok2020capitalfm}.
These hashtags have no actual meaning, but are thought to be part of a TikTok curation algorithm, which creator wishes to enhance a video's exposure in the ``For You'' list, which is where users spend the majority of their time, and that visibility may propel the video to immediate viral fame.
The other type are the trending hashtags, which are displayed on the {\em Discover} page and appear in our hashtag collection.
We consider the creators who ``hijack'' such hashtags to be looking for a boost in the visibility of their video.

\begin{table}
\centering
\small
  \begin{tabular}{lr| lr}
    \toprule
  \textbf{Hashtag} & \textbf{Frequency}&\textbf{Hashtag} & \textbf{\#Views}\\
    \midrule
    \#fyp* &20K & \#fyp &9.2T\\
    \#foryou* &8.5K & \#foryou &7.4T\\
    \#foryoupage* &6K & \#foryoupage &5.2T\\
    \#defrosting &5K & \#viral &2.7T\\
    \#homebusiness &4.6K & \#tiktok &1.5T\\
    \#sidetable &4.6K &\#funny &857B\\
    \#viral* &4.5K&\#tiktokindia &814B\\
    \#90saesthetic &4K&\#comedy &792B\\
    \#colorblast &4K&\#foryourpage &572B\\
    \#psychoedelicclown &3.8K&\#love &433B\\
    \#flipcard &3.7K&\#like &272B\\
    \#competitvegaming &3.7K&\#dance &267B\\
    \#ufosky &2.6K& \#viralvideo &237B\\
    \#parati* &2.4K&\#featureme &233B\\
  \bottomrule
\end{tabular}
  \caption{The left column presents the most frequent hashtags appear in our dataset as of March 2021. 
  Hashtags with an * are generic popular hashtags on TikTok, and do not appear in our hashtag collection. 
  The right column shows the most viewed hashtags on TikTok according to a report of the most popular hashtags in 2021~\cite{mau2021popularhashtags}.} 
  \label{tbl:termfrequency}
\end{table} 

\subsection{RH 3: Creator's Profile}
We posit that particularly popular accounts are more likely to see their content go viral.
This RH, which represents the creator's relationship with the platform and other TikTok users, includes the number of followers and their verified status.

\descr{F.9 Verified creator.}
A blue checkmark next to the account's name indicates that TikTok has confirmed that the account belongs to the user it represents~\cite{tiktok2019verified}.
Celebrities, non-profits, or official brand pages are often verified creators on TikTok.
According to TikTok, several factors are taken into account to grant a verified badge: authenticity, uniqueness, and engagement are mentioned in the official guidelines~\cite{tiktok2019verified}.

\descr{F.10 Popular creator.}
Previous research~\cite{lim2012following,haenlein2020navigating} on TikTok, Instagram, and Twitter used the number of followers and engagement rates as a starting point to identify a Popular creator (for example, at least 10,000 followers and an engagement rate above 10\%, which means at least 1,000 {\em likes}, comments, or shares per post on average for a Popular creator with 10,000 followers).
Based on the distribution of followers and high engagement of the TikTok platform, we define a Popular creator as a creator with at least 10,000 followers.
A Popular creator is not always a verified creator.

\section{Labeling}
\label{sec:annotation}
While some of the features from our codebook can be processed automatically (specifically, F.7 Lifespan, F.8 Hashtag, F.9 verified creator, and F.10 Popular creator), others are more nuanced and require human labeling.
Therefore, we manually code the remaining features: F.1 Subject, F.2 Scale, F.3 Point of view, F.4, Text, F.5 Emotion Experience, and F.6 style.
Although computer vision research~\cite{wu2015deep, kavukcuoglu2010learning, parizi2012reconfigurable} could in theory help identify subjects in the video, these features are highly subjective; overall, it would be difficult for an automated approach to label them correctly and consistently.
All features are coded in a binary way, making it easier for the annotators to agree on, except for F.2 Scale and F.3 Point of view, which include three exclusionary codes.

The labeling was conducted by three authors of this paper who have extensive experience with online communities in general and memetic remixes.
A subset of 20 randomly selected videos was used to calculate agreement, more precisely, 10 viral videos and 10 non-viral ones from different hashtag searches.

We reach a generally high score in agreement, as reported in Table~\ref{tbl:agreement}.
The table reports Fleiss' Kappa score, which ranges from 0 to 1 (0 indicates no agreement and 1 perfect agreement). 
Four of the six labels have an almost perfect agreement (score above 0.8).
Only two labels fall below the threshold of perfect agreement and have substantial agreement (score between 0.6 and 0.8); specifically, F.5 Emotion Experience, and  F.6 Style.
Although our codebook has a definition for it, the perception of emotion is quite subjective, and different annotators can label the same video as emotional or not, and this reduces the agreement between annotators.
Similarly, different perceptions of a video's style compared to other videos with the same hashtag create disagreements.
After establishing that the features in the codebook achieved high agreement between expert annotators, the first author annotated the rest of the videos.

\begin{table}
\centering
\small
  \begin{tabular}{lr}
    \toprule
  \textbf{Feature} & \textbf{Kappa}\\
    \midrule
     F.1 Subject & 0.89\\
     F.2 Scale & 0.89\\
     F.3 Point of View & 0.81\\
     F.4 Text & 0.89 \\
     F.5 Emotion Experience &0.69 \\
     F.6 style & 0.63\\
  \bottomrule
\end{tabular}
  \caption{Agreement between annotators.} 
  \label{tbl:agreement}
\end{table} 

\section{Modeling Indicators of (Short Videos) Virality}
\label{sec:model}

After identifying potential indicators of virality in short videos, we are interested in understanding if the features in our codebook are indeed able to discriminate between viral and non-viral TikTok videos.
To this end, we train several classifiers and find that the trained models can achieve high performance (up to 0.93 AUC).
We then analyze the most important features in our models, with the goal of understanding which elements in a video contribute the most to its virality.
Finally, we investigate whether the models trained on our label dataset generalize to other types of TikTok videos, like the ones posted on Twitter or that appear on the {\em Home} page as selected {\em ``For You'' } videos.

\subsection{Classification}
To build the machine learning models, we use the ten features described in Section~\ref{sec:characterizing}.
We experiment with five different classifiers: Random Forest~\cite{liaw2002classification}, Support Vector Machines (SVM)~\cite{suykens1999least}, Logistic Regression~\cite{hosmer2013applied}, Gaussian Bayesian~\cite{williams1998bayesian}, and Decision Trees~\cite{safavian1991survey}.
 
For each classifier, we take the annotated set of 400 short videos and perform 10-fold cross-validation; i.e., we randomly divide the dataset into ten sets and use nine for training and one for testing.
We repeat this process 10 times and calculate the average Area Under the ROC Curve (AUC) as well as precision, recall, and F1-score. 
The results are reported in Table~\ref{tbl:allresults}.
Overall, all classifiers achieve good results on our labeled dataset, with Logistic Regression achieving the best performance (AUC=0.93).

\begin{table}[t]
\small
\centering
 \setlength{\tabcolsep}{2pt}
  \begin{tabular}{lrrrrr}
    \toprule
    \bf{Classifier} & \bf{AUC} &\bf{F1-score}& \bf{Precision} & \bf{Recall} & \bf{Accuracy}\\
    \midrule
     Logistic Regression & 0.93 & 0.88 & 0.88 & 0.88 & 0.83\\
     SVM & 0.92 & 0.90 & 0.90 & 0.90 & 0.83\\
     Decision tree & 0.90 & 0.88 & 0.88 & 0.88 & 0.84\\
     Gaussian Bayesian & 0.89 & 0.80 & 0.80 & 0.80 & 0.77\\
     Random Forest & 0.89 & 0.85 & 0.85 & 0.85 & 0.79 \\
     \midrule
    Decision Tree (RH1) & 0.81 & 0.65 & 0.78 & 0.78 & 0.78 \\
    Random Forest (RH2) & 0.71 & 0.56 & 0.68 & 0.68 & 0.68\\
    Random Forest (RH3) & 0.86 & 0.83 & 0.85 & 0.85 &0.85\\
  \bottomrule
  \end{tabular}
  \caption{Performance of all classifiers and best performance classifiers for each Research Hypothesis.}
  \label{tbl:allresults}
\end{table}

We also look at the impact of each RH in isolation; see the last three rows in Table~\ref{tbl:allresults}.
The features in RH3 (Creator's profile) have the best performance. 
The Random Forest classifier achieves an AUC of 0.86, suggesting that the popularity of a user on TikTok has an impact on their video's virality.
The features in RH1 (Content element) also lead to good prediction.
Decision Tree yields an AUC of 0.81, confirming that the content of a video can influence a video's virality.
The features in RH2 (Recommendation System) have the lowest AUC among the three RHs, as low as 0.71.
In fact, the accuracy obtained on these features is also quite low (0.56), suggesting that they are not a good predictor of a video's virality.

\subsection{Feature Analysis} 
\label{sec:feature_analysis}

Next, we analyze what features contribute the most to determine if a TikTok video will go viral or not.
To this end, we perform a feature analysis of the best-performing classifier (Logistic Regression) to identify indicators of virality (or non-virality). 
The top five features discovered by this model are discussed below.

\begin{enumerate}

\item \textbf{\em Popular creator:} 
The most important feature is whether or not its creator is popular. 
Short videos uploaded by a creator with 10,000 or more followers have a higher chance of going viral. 
This trait is found in 90\% of viral short videos in our dataset, but just 24\% of non-viral ones.

\item \textbf{\em Scale:} 
The second most significant element is the scale of the video.
While a close-up or a medium-shot scale does not seem to affect virality in either direction (with 64\% and 54\% of the viral and non-viral videos in our dataset presenting a close-up scale, 35\% of the viral videos in our dataset present this feature, while only 29\% of the non-viral ones), we find that videos that use a large shot scale are less likely to be viral (17\% of the non-viral memes present this feature, while only 1\% of the viral ones).\
This matches previous studies on image memes~\cite{ling2021dissecting} suggesting that highly viral memes are more likely to use a close-up or medium-shot scale.

\item \textbf{\em Lifespan:}
  The Logistic Regression classifier recognizes the lifespan in which a video stays on the platform as a key attribute.
Though the majority of TikTok videos are created recently, 98\% of viral short videos are at the prime of their lifespan, while 85\% of non-viral ones are too old to be seen.

\item \textbf{\em Text:} 
Whether or not the video contains text also has an impact on its virality. 
In comparison to 67\% of viral videos, 30\% of non-viral videos have text to guide viewers.
This is in contrast with the conclusion of previous studies, which found that text impedes virality in image memes~\cite{ling2021dissecting}.

\item \textbf{\em Point of View:} 
  We find that the point of view of a video impacts its chances to go viral.
70\% of viral short videos are shot in the second point of view, whereas only 21\% are shot in the first point of view; for non-viral short videos, these percentages are 60\% and 28\%, respectively.
\end{enumerate}

\subsection{Short Videos on Twitter and \#fyp.} 

Our findings show that our codebook can characterize indicators of virality on a dataset of trending short videos that appeared on TikTok's {\em Discover} page. 
A key question is whether these results generalize beyond the {\em Discover} page to other videos posted on the platform.
To answer this question, we look at whether the indicators discovered by our model can predict the virality of random TikTok videos uploaded on Twitter and the most popular TikTok videos.

To this end, we collect TikTok videos from two sources.
First, we collect 50 TikTok videos posted on Twitter between March 11th and April 11th, 2021.
These videos are randomly sampled from the TikTok links that appeared in the 1\% Twitter Streaming API during that period.
Next, we collect the top 20 viral short videos that appeared when querying the \#fyp hashtag on April 11, 2021.
According to TikTok, this hashtag drew over 10 trillion views in total, see Table~\ref{tbl:termfrequency}.
\#fyp is an abbreviation for ``For Your Page,'' which refers to the personalized {\em Home} page (based on user activity).
We then apply the same criterion discussed in Section~\ref{sec:dataset} to determine the virality of these videos.
In total, we have 48 viral videos and 22 non-viral videos.

We then follow the same annotation process that we used for our ground truth of 400 videos on these 70 videos.
Note that we are unable to annotate F.8 Style based on Twitter-sourced TikTok videos because, as discussed in Section~\ref{sec:annotation}, we need to compare the video with the other videos with the same hashtag. 
Finally, we run the classifiers trained on our ground truth on this new dataset of 70 videos.
Our classifiers correctly predict all viral and non-viral TikTok short videos. %
This suggests that our indicators can generalize beyond our training dataset and can indeed characterize the virality of any TikTok video.

\section{Case Studies}

In this section we provide an analysis of a viral and non-viral TikTok video identified by our approach.

\subsection{A viral video example}
On March 12th, 2021, a popular pet account with 6.9 million followers posted a brief video of the owner and her cat.
In the video, a lady raises her cat, facing the camera in a second-person view, and tries to teach her cat to say syllables like ``Mom.''
This video received 6.8 million {\em likes} on TikTok, making it viral. 
It is shot in the second person perspective, with the character's face tightly framed in a close-up scale.
The creator added text on the video to guide viewers to follow the conversation.
This video was created during the period when some of the hashtags attached with the video were promoted on the ``Discover'' page, i.e., \#AthletesofTikTok, \#Defrosting, etc.
Other popular generic hashtags, like ``\#fyp,''``\#foryou,'' ``\#foryoupage,'' ``\#trending,'' are included, possibly to increase visibility of the video.
Although the account has a large number of followers, the account is unverified by TikTok.
However, this large follower base constitutes a substantial potential reach fr the video.
At the time we collected this video, it was ranked first under the hashtag event \#AthletesofTikTok. %

This viral video meets all the top five features in predicting a video's virality as discussed in Section~\ref{sec:feature_analysis}.
First, the video is created by a popular creator with 6.9 million followers.
Second, the subject of the video is tightly framed in the video in a small scale shot.
Third, the video includes several hashtags while those hashtags were trending, including \#AthletesofTikTok and \#Defrosting. %
Fourth, the creator added text on the video to guide viewers to understand what she is saying to her cat.
Fifth, the video is in a second point of view, where the creator and her cat are facing the camera, talking to the viewers.
Our approach correctly models these indicators of virality.

\subsection{An non-viral video example}

A 15-second video of defrosting a truck posted on February 12th, 2019 received 28 likes over two years until we collected its metadata. %
The video was created by an account with 1,799 followers.
In comparison to popular accounts with over 10,000 followers, this account has a limited impact and cannot reach a large number of TikTok users.
The creator put several related hashtags in the description, for example ``\#ifyourcoldtherecold,'' ``\#defrosting,'' ``\#lovemyheatedshop,'' etc.
None of these hashtags were promoted on the ``Discover'' page at the time the video was published.
In the video, the creator shot his truck from the rear to the front; more precisely, on a chilly winter day, the video shows the creator's truck while defrosting.
The video is only 15 seconds long and the video quality is low.
The camera moves in such a quick way that viewers can hardly get a good look at the truck on the screen.
No text is guiding the viewers to understand the video's subject, too.

This video is categorized as a non-viral video for receiving only a small number of {\em likes}.
It does not possess four out of five top features identified by our model, which help promote virality.
First, the creator of this video does not have enough followers to qualify as a popular creator.
Second, this video was published in 2019, and none of its hashtags were trending at that time. 
Third, there is no text attached to lead viewers to follow the video's subject.
Though this video is shot on a medium scale from the first point of view, the shaking camera makes the video very vague.
Our model correctly predicts this video as non-viral.

\section{Discussion \& Conclusion}
\label{sec:discussion}

In this paper, we studied what elements in short videos posted on TikTok contribute to their virality.
We focused on different elements of the videos, based both on previous studies in viral content and TikTok's platform design, and identified several elements as features that may impact the video's virality.

We focused on the videos that received the most and least {\em likes} from the result of the hashtag search, and used the {\em likes} that a video receives as a proxy for its virality. %
We then relied on a mixed-method approach to develop a codebook that depicts the indicators of virality in short videos, and trained classifiers to investigate whether these indicators can successfully predict if a video will go viral. 
The best performance classifier, Logistic Regression, achieves an AUC of 0.93, indicating that it can successfully distinguish between viral and non-viral videos.
We also generalized our model to the \#fyp hashtag and short videos on Twitter, and showed that our classifiers correctly predict all videos.

Overall, our results show that the number of followers is the most reliable indicator of a video's virality; a creator with over 10,000 followers is more likely to make a viral video than the others.
Moreover, close-up or medium-shot videos using a second point of view are more common among viral videos.
In contrast to previous studies on image memes, text does not impede video's virality but rather facilitates it. 
Finally, videos created recently are more likely to go viral.

\descr{Design Implications.} Our work provides a first look at what indicators contribute to the virality of short videos on TikTok.
These findings have clear implications with respect to marketing and campaigning and could be used by activists or advertisers to deliver more effective messaging.
In recent years, traditional marketing methods are being phased out in favor of creator-centric strategies; not only on TikTok, but on other media-oriented services like Instagram and Youtube as well. 
Our work sheds light on the indicators of virality in TikTok videos. 
It should come as no surprise that a creator's number of followers is the most important factor in making a video viral. 
However, how a video is shot also influences its chances of going viral.
From the second point of view, it is easier to get a clean, steady shot. 

We also found that emotion has little bearing on the virality of a video, which is consistent with marketing research: basic emotions (sadness, anger, surprise, fear, and disgust) do not predict the viewer's reaction to an ad or its effectiveness~\cite{lewinski2014predicting}.
Another interesting finding is that common practices thought to improve the chances of content going viral, like hasthag stuffing, do not seem to have much of an effect on virality on TikTok.

Alas, the observations made in this paper could be used by malicious actors to make unwanted content become more convincing and spread further (e.g., spam, fraud, conspiracies). 
On the other hand, online services like TikTok could also use our insights to develop more effective moderation techniques.
For example, an automated system could rank videos by their likelihood of going viral, and allow TikTok to direct moderation resources to these videos first.
After all, the videos that are likely to reach a wider audience are the ones that have the potential of causing the most harm, especially when dealing with fraud or misinformation.

As for hashtag stuffing, which is often used by content creators to increase the popularity of their videos, we note that simple heuristics could be used to detect and limit this phenomenon, even though it does not seem to have a particular effect on virality.

\descr{Limitations \& Future Work.} Naturally, our work is not without limitations.
First, since TikTok does not expose an API to allow researchers to access their content, we had to rely on their search functionality to identify suitable videos for our study.
This might potentially lead to a bias, in the sense that mostly popular videos are returned, in particular in relation to popular hashtags.
Since TikTok does not release details on how their recommendation algorithm works, it is not possible to quantify this bias.

Second, while we identify viral and non-viral videos based on thresholds of their number of likes, virality is a spectrum. 
While our results work well in characterizing videos that go viral from those almost never shared, they may not be as clearcut in determining indicators of virality for videos that somewhat fall in the middle.

In future work, we plan to investigate how users {\em engage} with viral and non viral videos.
For example, how are viral videos discussed on social media when they are re-shared? 
What kind of comments do viral and non-viral videos attract?
Can we predict what videos will receive positive comments, and which ones will be targeted by hate speech?
Answering these questions will require both advancements in the theoretical framework developed for this paper as well as overcoming the engineering challenges involved with automated large-scale data collection on TikTok.

\descr{Acknowledgments.}
This work was supported by the National Science Foundation under grants CNS-1942610, IIS-2046590, CNS-2114407, and CNS-2114411, and by the Hariri Institute for Computing at Boston University.

\small
\bibliographystyle{abbrv}

\end{document}